\documentclass[pra,aps,showpacs,epsf,superscriptaddress,twocolumn]{revtex4}

\def\be{\begin{equation}}
\def\ee{\end{equation}}

\usepackage{bm}
\usepackage{amsmath}
\usepackage{amssymb}
\usepackage[pdftex]{graphicx}
\usepackage{color}
\usepackage[usenames,dvipsnames]{xcolor}
\usepackage{mathrsfs}
\usepackage[toc,page]{appendix}
\usepackage[abs]{overpic}
\usepackage{array}

\newcommand{\bea}{\begin{eqnarray}}
\newcommand{\eea}{\end{eqnarray}}
\newcommand{\bi}{\begin{itemize}}
\newcommand{\ei}{\end{itemize}}

\newcommand{\oo}{\Omega_1}
\newcommand{\ot}{\Omega_2}
\newcommand{\dt}{\frac{d}{dt}}
\newcommand{\ddt}[1]{\frac{d#1}{dt}}
\newcommand{\dz}{\frac{\partial}{\partial z}}
\newcommand{\ddz}[1]{\frac{\partial #1}{\partial z}}
\newcommand{\vel}{ v}
\newcommand{\etal}{{\it{et~al.}}}
\newcommand{\ket}[1]{\left| #1 \right>} 
\newcommand{\bra}[1]{\left< #1 \right|} 
\newcommand{\for}{\mathcal F}
\newcommand{\dif}{\mathcal D}

\newcommand{\isat}{I_{\rm{sat}}}
\newcommand{\delt}{\delta^{\left|\mbox{\tiny $2'$}\right>}}
\newcommand{\delth}{\delta^{\left|\mbox{\tiny $3'$}\right>}}
\newcommand{\gamt}{\Gamma^{\left|\mbox{\tiny $2'$}\right>}}
\newcommand{\gamth}{\Gamma^{\left|\mbox{\tiny $3'$}\right>}}
\graphicspath{{../images/}}

\begin{document}

\title{$\Lambda$-enhanced Sub-Doppler Cooling of Lithium Atoms in $D_1$ Gray Molasses}

\author{Andrew T. Grier}
\affiliation{Laboratoire Kastler-Brossel, \'Ecole Normale Sup\'erieure, CNRS and UPMC, 24 rue Lhomond, 75005 Paris, France}
\author{Igor Ferrier-Barbut}
\affiliation{Laboratoire Kastler-Brossel, \'Ecole Normale Sup\'erieure, CNRS and UPMC, 24 rue Lhomond, 75005 Paris, France}
\author{Benno S. Rem}
\affiliation{Laboratoire Kastler-Brossel, \'Ecole Normale Sup\'erieure, CNRS and UPMC, 24 rue Lhomond, 75005 Paris, France}
\author{Marion Delehaye}
\affiliation{Laboratoire Kastler-Brossel, \'Ecole Normale Sup\'erieure, CNRS and UPMC, 24 rue Lhomond, 75005 Paris, France}
\author{Lev Khaykovich}
\affiliation{Department of Physics, Bar-Ilan University, Ramat-Gan, 52900 Israel}
\author{Fr\'ed\'eric Chevy}
\affiliation{Laboratoire Kastler-Brossel, \'Ecole Normale Sup\'erieure, CNRS and UPMC, 24 rue Lhomond, 75005 Paris, France}
\author{Christophe Salomon}
\affiliation{Laboratoire Kastler-Brossel, \'Ecole Normale Sup\'erieure, CNRS and UPMC, 24 rue Lhomond, 75005 Paris, France}
\date{\today}

\begin{abstract}
Following the bichromatic sub-Doppler  cooling scheme on the $D_{1}$ line of $^{40}$K recently demonstrated in D. R. Fernandes \emph{et al.} (2012), we introduce a similar technique for $^{7}$Li atoms and obtain temperatures of $60\,\mu {\rm K}$ while capturing all of the $5\times10^{8}$ atoms present from the previous stage. We investigate the influence of the detuning between the the two cooling frequencies and  observe a threefold  decrease of the temperature when the Raman condition is fulfilled. We interpret this effect as arising from extra cooling due to long-lived coherences between hyperfine states. Solving the optical Bloch equations for a simplified, $\Lambda$-type three-level system we identify the presence of an efficient cooling force near the Raman condition. After transfer into a quadrupole magnetic trap, we measure a phase space density of $\sim 10^{-5}$. This laser cooling offers a promising route for fast evaporation of lithium atoms to quantum degeneracy in optical or magnetic traps.

\end{abstract}

 \maketitle

\section*{Introduction}

Lithium is enjoying widespread popularity in the cold-atom trapping community thanks to the tunability of its two-body interactions and its lightness.
Both the fermionic and bosonic isotopes of lithium feature broad, magnetically-tunable Feshbach resonances in a number of hyperfine
states~\cite{chin2010}.  The presence of these broad resonances makes lithium an attractive candidate for studies of both the Fermi- and Bose-Hubbard models~\cite{blochDalibardZwerger} and the strongly correlated regime for bulk dilute gases of Fermi~\cite{nascimbene2010} or Bose~\cite{Navon:2011dz, Wild:2012fi, Rem:2013fg} character. Its small mass and correspondingly large photon-recoil energy are favorable factors for large area atom interferometers \cite{vigue2012} and precision frequency measurements of the recoil energy and fine structure constant \cite{Bouchendira:2011cc}.  Under the tight-binding lattice model, lithium's large photon-recoil energy leads to a larger tunneling rate and faster timescale for super-exchange processes, allowing for easier access to spin-dominated regimes
~\cite{esslinger2010}.
Finally, lithium's small mass reduces the heating due to non-adiabatic parts of the collision between ultracold atoms and Paul-trapped ions.  This feature, together with Pauli suppression of atom-ion three-body recombination events involving $^{6}$Li~\cite{harter2012}, potentially allows one to reach the $s$-wave regime of ion-atom collisions~\cite{cetina2012}.

However, lithium, like potassium, is harder to cool using optical transitions than the other alkali atoms.  The excited-state structure of the $D_{2}$ transition in lithium lacks the separation between hyperfine states  for standard sub-Doppler cooling techniques such as polarization gradient cooling~\cite{dalibard1989,lett1989,weiss1989} to work efficiently.  Recently, it has been shown by the Rice group that cooling on the narrow $2S_{1/2}\to 3P_{3/2}$ transition  produces lithium clouds near $60\,\mu$K, about half the $D_2$-line Doppler cooling limit~\cite{HuletUV}, and can be used for fast all-optical production of  a $^{6}$Li quantum degenerate Fermi gas.  However, this approach requires special optics and a coherent source at 323 nm, a wavelength range where power is still limited. Another route is to use the three-level structure of the atom as implemented previously in neutral atoms and trapped ions \cite{gupta1993,zerussians,drewsen1995,roos2000,Grynberg:1994, morigi2007,dunn2007}. The three-level structure offers the possibility of using dark states to achieve temperatures below the standard Doppler limit, as evidenced by the use of velocity-selective coherent population trapping (VSCPT) to produce atomic clouds with sub-recoil temperatures \cite{Aspect:1988fy}. In another application, electromagnetically induced transparency has been used to demonstrate robust cooling of a single ion to its motional ground state~\cite{roos2000,Morigi:2000in}.

\begin{figure}
\centerline{\includegraphics[width=0.6\columnwidth]{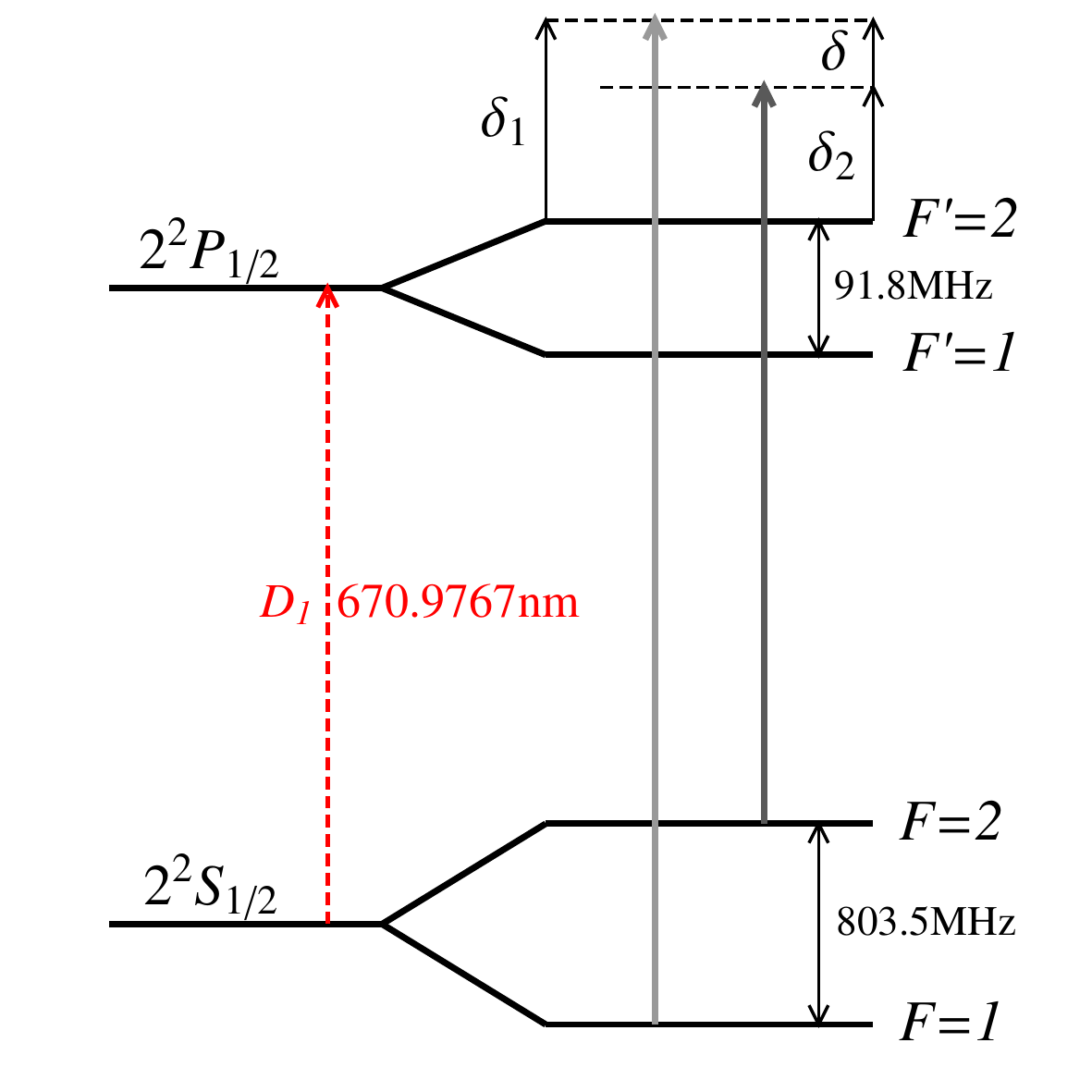}}
\caption{(Color online) The $D_1$-line for $^{7}$Li. The cooling scheme has a strong coupling laser (principal beam, black solid arrow) $\delta_2$ blue detuned from the $\ket{F=2}\rightarrow \ket{F'=2}$ transition and a weak coupling laser (repumper, gray solid arrow) $\delta_1$ blue detuned from the $\ket{F=1}\rightarrow \ket{F'=2}$ transition. The repumper is generated from the principal beam by an electro-optical modulator operating at a frequency $803.5\,{+}\,\delta/2\pi\,{\rm MHz}$ where $\delta=\delta_1-\delta_2$. }
\label{fig:levels}
\end{figure}

In this paper, we implement three-dimensional bichromatic sub-Doppler laser cooling of  $^{7}$Li atoms on the $D_{1}$ transition.  Fig.\ref{fig:levels} presents the $^{7}$Li level scheme and the detunings of the two cooling lasers that are applied to the atoms after the magneto-optical trapping phase. Our method combines a gray molasses cooling scheme on the $\ket{F=2}\rightarrow \ket{F'=2}$ transition~cite{weidemuller1994,fernandes2012 } with phase-coherent addressing of the $\ket{F=1}\rightarrow \ket{F'=2}$ transition, creating VSCPT-like dark states at the two-photon resonance. 
  Instead of UV laser sources, the method uses laser light that is conveniently produced at 671~nm by semi-conductor laser sources or solid-state lasers \cite{eismann2010laser, Eismann:wc} with sufficient power. This enables us to capture all of the  $\simeq5\times10^{8}$ atoms from a MOT and cool them to $60\,\mu$K in a duration of 2\,ms.

We investigate the influence of the relative detuning between the two cooling lasers and  observe a threefold  decrease of the temperature  in a narrow frequency range around the exact Raman condition. We show that extra cooling arises due to long-lived coherences between hyperfine states.
We develop a simple theoretical model for a sub-Doppler cooling mechanism which occurs in atoms with a $\Lambda$-type three-level structure, in this case, the $F=1$, $F=2$, and $F^{\prime}=2$ manifolds of the $D_{1}$ transition in $^{7}$Li.  
The main physical cooling mechanism is contained in a 1D bichromatic lattice model.  We first give a perturbative solution to the model and then verify the validity of this approach with a continued fraction solution to the optical Bloch equations (OBEs).

\section{Experiment}
\begin{figure}[htbp]
\begin{centering}
\begin{overpic}[width=.95\columnwidth]{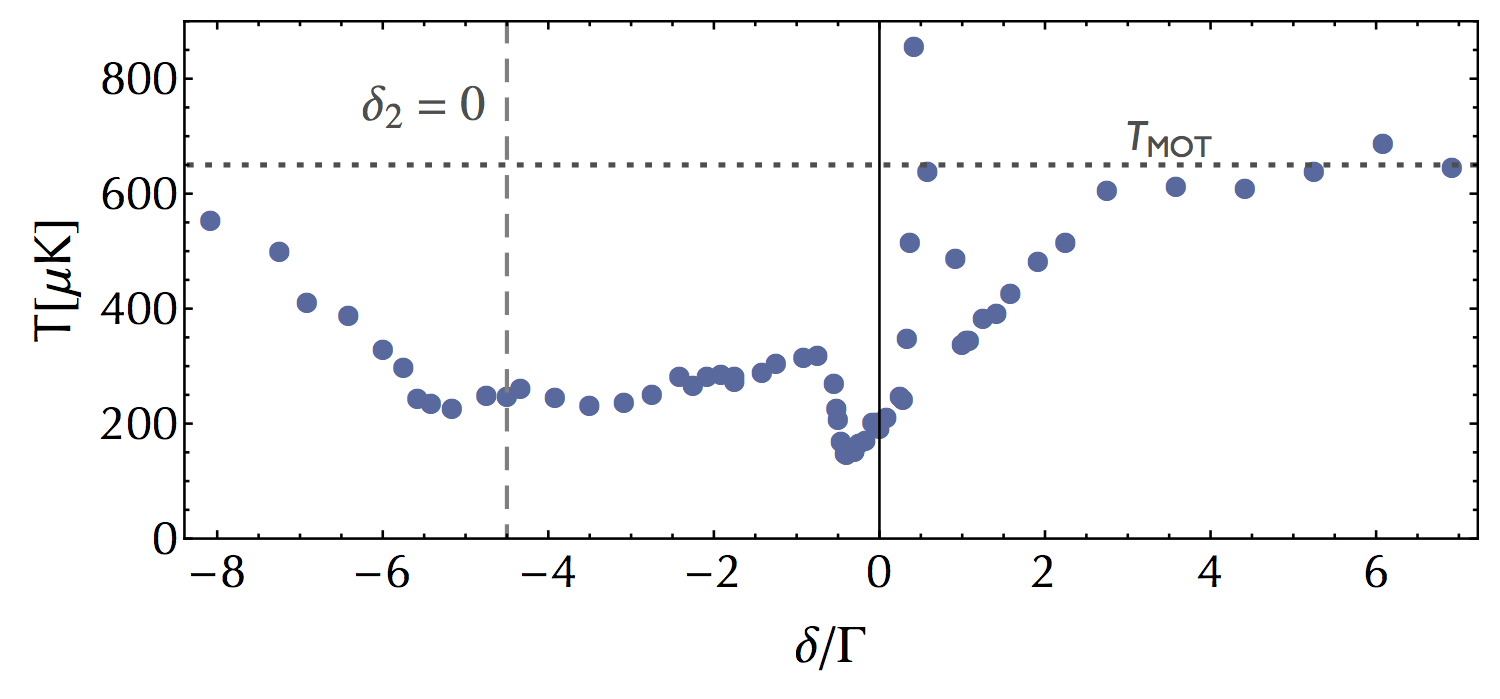}
\put(70,15){(a)}\end{overpic}\\
\begin{tabular}{cc}
\begin{overpic}[width=.45\columnwidth]{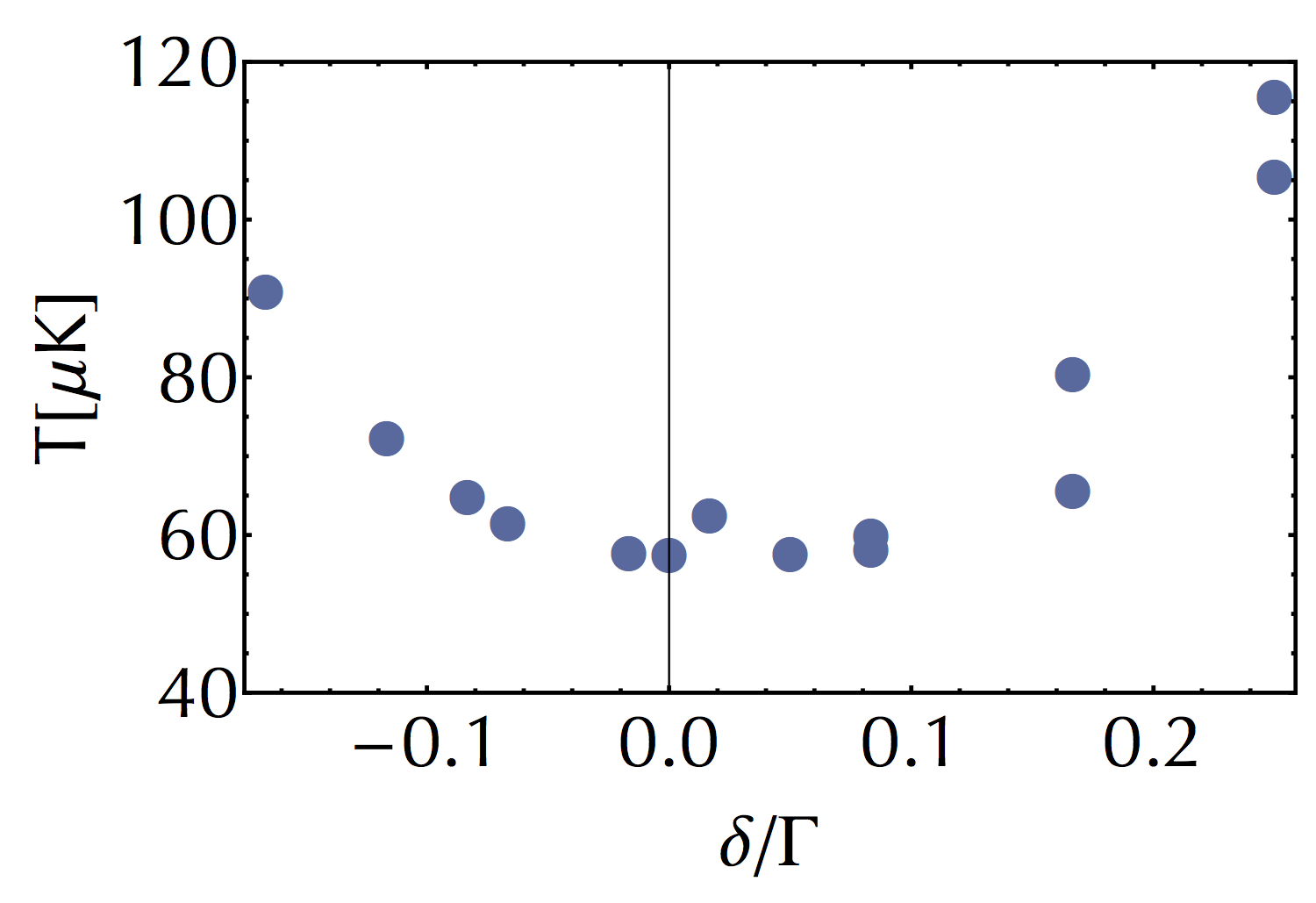} \put(10,20){(b)}\end{overpic}&
\begin{overpic}[width=.45\columnwidth]{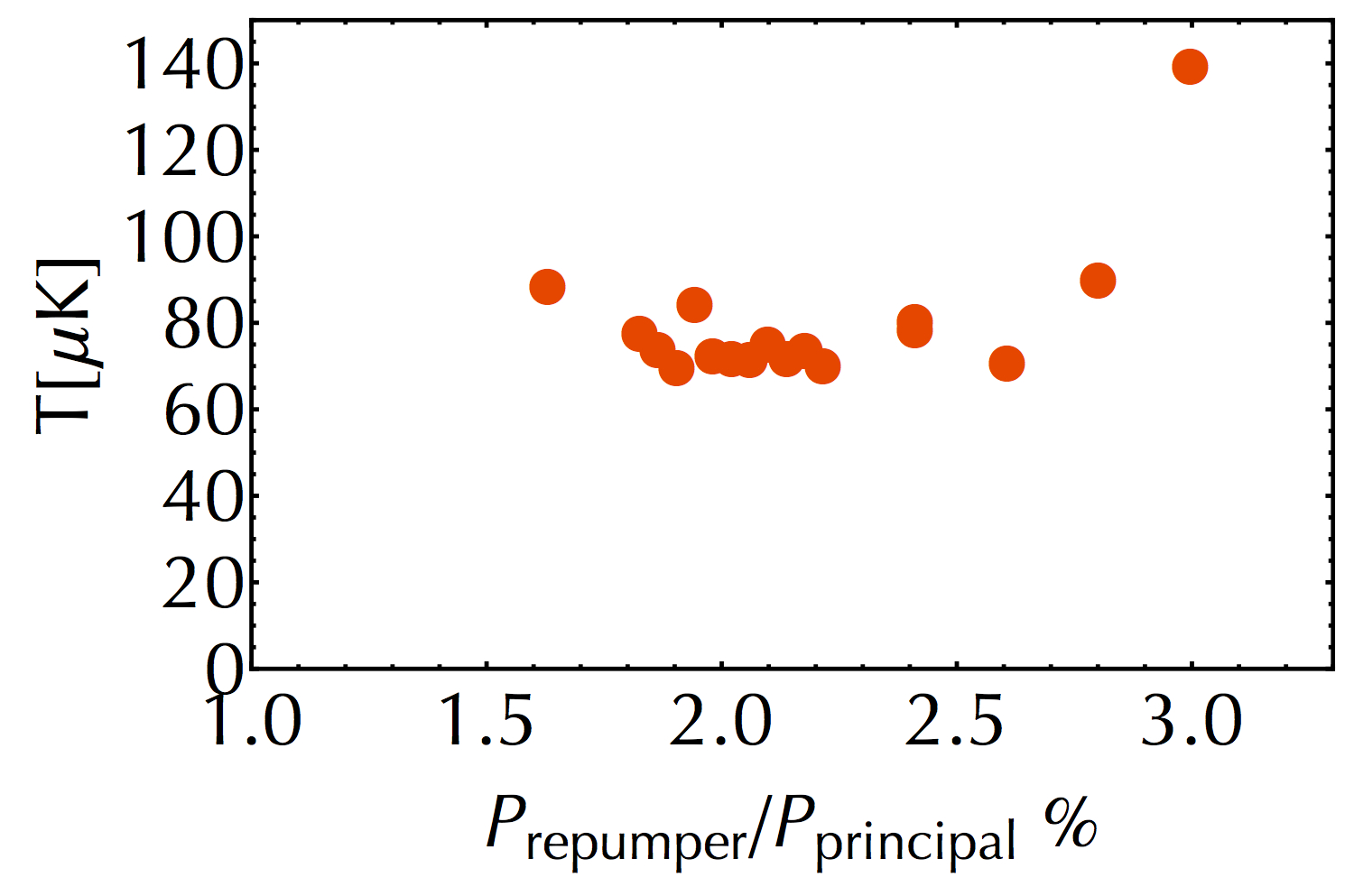} \put(30,10){(c)}\end{overpic}\\
\end{tabular}
\caption{(Color online) a)Typical temperature of the cloud as a function of the repumper detuning for a fixed principal beam detuned at $\delta_1 = 4.5 \Gamma=2\pi\times26.4\,$ MHz. The dashed vertical line indicates the position of the resonance with transition $\ket{F=2}\rightarrow\ket{F'=2}$, the dotted horizontal line shows the typical temperature of a MOT. b) Zoom-in on the region near the Raman condition with well-aligned cooling beams and zeroed magnetic offset fields.  c) Minimum cloud temperature as a function of repumper power.}
\label{fig:data0}
\end{centering}
\end{figure}


The stage preceding $D_{1}$ sub-Doppler cooling is a compressed magneto-optical trap (CMOT) in which, starting from a standard MOT optimized for total atom number, the frequency of the cooling laser is quickly brought close to resonance while the repumping laser intensity is diminished in order to increase the sample's phase space density~\cite{mewes1999}.  The CMOT delivers $5 \times10^{8}$ $^7$Li atoms at a temperature of 600~$\mu{\rm K}$.  The atoms are distributed throughout the $F=1$ manifold in a spatial volume of 800-$\mu$m $1/e$ width.  Before starting our $D_1$ molasses cooling, we wait 200~$\mu$s to allow any transient magnetic fields to decay to below 0.1~G.
The light used for $D_1$ cooling  is generated by a solid-state laser presented in \cite{eismann2010laser}. 
The laser is locked at frequency $\omega_2$, detuned from the $\ket{F=2}\rightarrow \ket{F'=2}$ $D_{1}$ transition in $^{7}$Li by $\delta_2$. It is then sent through a resonant electro optical modulator (EOM) operating at a frequency near the hyperfine splitting in $^{7}$Li, $\nu_{\rm EOM}=803.5\,{\rm MHz}\,{+}\,\delta/2\pi$. This generates a small amplitude sideband, typically a few \% of the carrier, at frequency $\omega_1$.  We define the detuning of this frequency from the $\ket{F=1}\rightarrow \ket{F'=2}$ transition as $\delta_1$ (such that $\delta=\delta_1-\delta_2$), as shown in Fig.~\ref{fig:levels}.  
Using about 150~mW of 671\,nm light we perform a three-dimensional $D_1$ molasses as in \cite{fernandes2012}, with 3 pairs of $\sigma^{+}-\sigma^{-}$ counter-propagating beams. The beams are of 3.4-mm waist and the intensity $(I)$ of each beam is $I \gtrsim 45\isat$ where $\isat= 2.54$\,mW/cm$^2$ is the saturation intensity of the $D_{2}$ cycling transition in Lithium.
We capture all of the atoms present after the CMOT stage into the  $D_{1}$ gray molasses.  The $1/e$ lifetime of atoms in the molasses is $\geq$50~ms. After being cooled for 1.5-2.0~ms, the temperature is a low as 40~$\mu$K without optical pumping or 60~$\mu$K after optical pumping into the $\ket{F=2,m_{F}=2}$ state for imaging and subsequent magnetic trapping.  In contrast with~\cite{fernandes2012}, we find no further reduction in the steady-state temperature by slowly lowering the light intensities after the initial 2.0~ms. 

During the molasses phase, we find a very weak dependence on the principal laser detuning for $3\Gamma\leq\delta_2\leq 6\Gamma$.  For the remainder of this article, we will use a principal laser detuning of $\delta_2=4.5\Gamma=2\pi\times 26.4$\,MHz. In Fig.~\ref{fig:data0}(a), the temperature dependence upon the repumper detuning is displayed for typical conditions. For  $-9\leq \delta/\Gamma \leq -6$, the temperature drops from 600~$\mu$K (the CMOT temperature) to  200~$\mu$K  as gray molasses cooling gains in efficiency when the weak repumper comes closer to resonance.
For $-6\leq \delta/\Gamma \leq -1$, the cloud temperature stays essentially constant but, in a narrow range near the position of the exact Raman condition ($\delta=0$), one notices a sharp drop of the temperature.  For $\delta$ slightly blue of the Raman condition, a strong heating of the cloud occurs, accompanied by a sharp decrease in the number of cooled atoms.  Finally for  $\delta \geq \Gamma$, the temperature drops again to a level much below the initial MOT temperature until the repumper detuning becomes too large to produce significant cooling below the CMOT temperature.

Figs.~\ref{fig:data0}(b) and (c) show the sensitivity of the temperature minimum to repumper deviation from the Raman condition and repumper power, respectively. The temperature reaches 60~$\mu$K in a $\pm 500\,$kHz interval around the Raman resonance condition.  After taking the data for Fig.~\ref{fig:data0}(a), the magnetic field zeroing and beam alignment were improved, which accounts for the frequency offset and higher temperature shown in (a) relative to (b) and (c).  The strong influence of the repumper around the Raman condition with a sudden change from cooling to heating for small and positive Raman detunings motivated the study of the bichromatic-lattice effects induced by the $\Lambda$-type level configuration which is presented in the next section.

\section{Model for hyperfine Raman coherence effects on the cooling efficiency}

\begin{figure}[hbp]
\begin{center}
\centerline{\includegraphics[width=0.95\columnwidth]{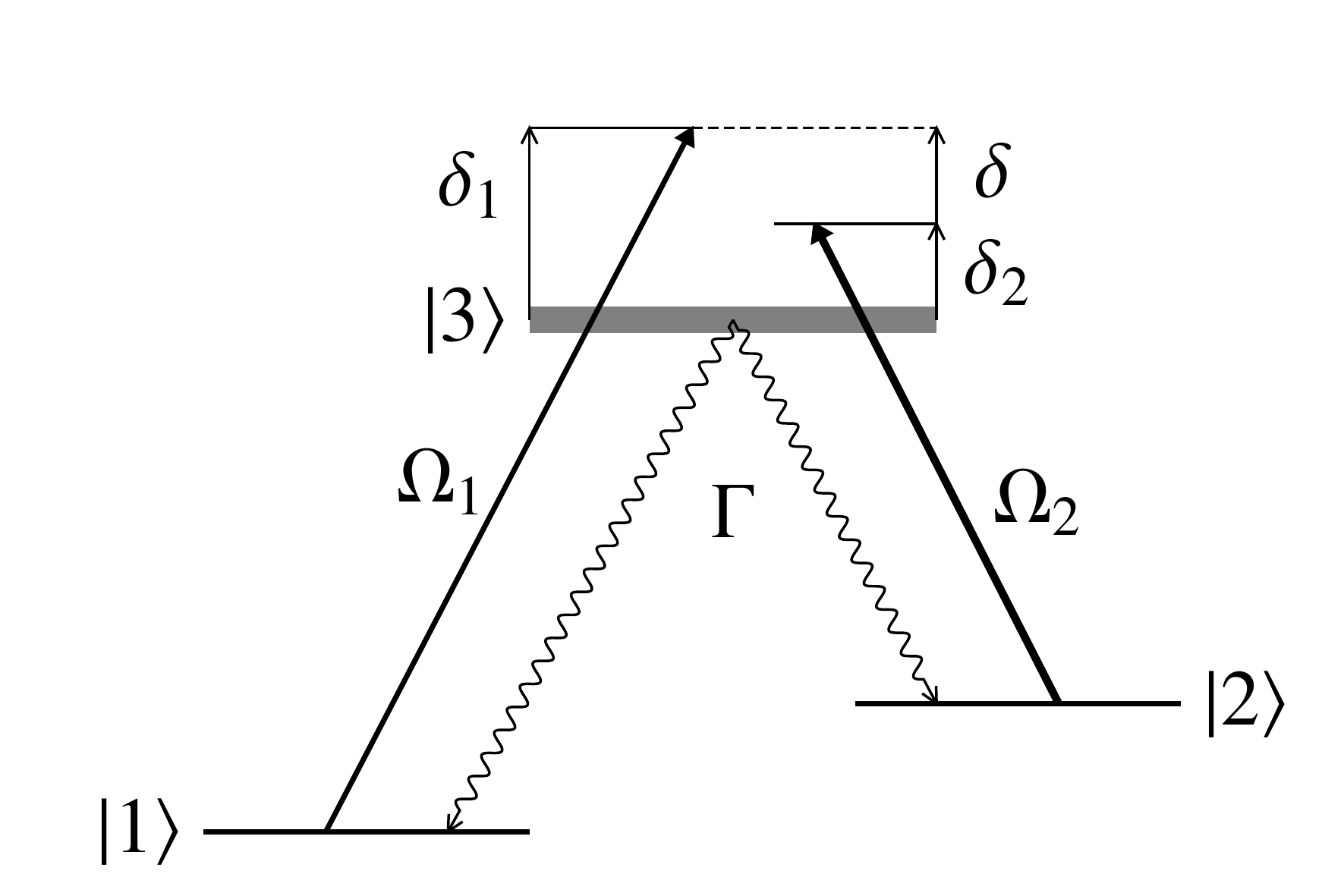}}
\caption{The $\Lambda$ level scheme. An intense  standing wave with Rabi frequency $\Omega_2$,  and a weaker standing wave with Rabi frequency $\Omega_1$, detuning $\delta_1$, illuminate an atom with three levels in a $\Lambda$ configuration.}
\label{fig:scheme}
\end{center}
\end{figure}

In order to understand how the addition of the second manifold of ground states modifies the gray molasses scheme, we analyze a one-dimensional model based on a $\Lambda$-type three-level system schematically represented in Fig.\ref{fig:scheme}.

\subsection{The model}

This model includes only the $F=1,2$ hyperfine ground states and the $F'=2$ excited state ignoring the Zeeman degeneracy, hence standard gray molasses cooling~\cite{weidemuller1994} will not appear in this model. The states are addressed by two standing waves with nearly the same frequency $\omega_1 \simeq \omega_2 \simeq \omega=kc$ but spatially shifted by a phase $\phi$. The principal cooling transition $F=2\rightarrow F'=2 $ is labeled here and below as transition 2, between states $\ket{2}$ and $\ket{3}$ with a Rabi frequency $\ot = \Gamma\sqrt{I/2I_{\rm{sat}}}$ where $I$ is the laser light intensity and $I_{sat}$ the saturation intensity on this transition. The repumper transition is labeled 1, between states $\ket{1}$ and $\ket{3}$ with Rabi frequency $\oo$ much smaller than $\ot$.

The corresponding Hamiltonian for the light-atom interaction in the rotating wave approximation (at $\omega$) is
\bea\label{eq:Val}
\begin{split}
\mathcal{\hat H_{\rm{a.l.}}}=&\,\hbar\ot \text{cos}(kz)\,(\ket{2}\bra{3}+ h.c.)\\&+\,\hbar\oo\, \text{cos}(kz+\phi)\, (\ket{1}\bra{3}+ h.c.)\,\\&+ \,\hbar\delta_2\ket{2}\bra{2}\,+\,\hbar\delta_1\ket{1}\bra{1}.
\end{split}
\eea


%

\noindent The usual formalism used to compute the atom's dynamics is to consider the light force as a Langevin force. Its mean value is $\for(\vel)$, and the fluctuations around this mean will give rise to diffusion in momentum space, characterized by the diffusion coefficient $\dif_p(v)\geq0$. In order to calculate an equilibrium temperature, one needs $\for(\vel)$ and $\dif_p(v)$. 
In the limit of small velocities the force reads 
\bea
\for(\vel)\simeq -\,\alpha\,\vel,
\eea
with $\alpha$ the friction coefficient. If $\alpha>0$ the force is a cooling force, in the opposite case it produces heating.
For a cooling force the limiting temperature in this regime is given by 
\bea\label{eq:limitT}
k_{{\rm B}}T\simeq\mathcal D_p(0)/\alpha.
\eea
However, since our model (\ref{eq:Val}) is a gross simplification of the physical system, we do not expect to be able to quantitatively predict  a steady-state temperature.  Instead, in order to reveal the physical mechanisms in action, we only calculate the force $\for(\vel)$ and the excited state population $\rho_{33}$. Restricting our analysis to the force and photon scattering rate, $\Gamma\cdot\rho_{33}$, suffices to determine whether the action of the weak repumper serves to heat or cool the atomic ensemble.

From (\ref{eq:Val}) the mean light force on the atoms is computed by taking the quantum average of the gradient of the potential, $F=\left< -\nabla\mathcal{\hat H_{\rm{a.l.}}} \right>\,=\,-\text{Tr}\left[\,\hat\rho\,\mathcal{\hat H_{\rm{a.l.}}}\right]$, with $\rho$ the density matrix, yielding the wavelength-averaged force $\for$
\bea
\hspace{-5pt}\for(\vel)&=&\frac{k}{2\pi}\int_0^{\frac{2\pi}{k}}dz\,F(z,\vel)
\eea
\bea\label{eq:avgForce}
\hspace{-5pt}\for(\vel)&=&\frac{\hbar k^2}{\pi}\int_0^{\frac{2\pi}{k}}dz \,{\rm{sin}}(kz)\left(\ot \text{Re}\rho_{23}+\oo \text{Re}\rho_{12} \right)
\eea
The spontaneous emission rate averaged over the standing wave is simply given by the linewidth of the excited state multiplied by its population:
\bea
\Gamma'=\frac{k}{2\pi}\int_0^{\frac{2\pi}{k}}\text dz\,\Gamma\,\rho_{33}.
\eea
\noindent So, both the force and the spontaneous emission rate are functions of the density matrix $\rho$, the evolution of which is given by the optical Bloch equations (OBEs)
\bea\label{eq:OBE}
i\dt\rho &=& \frac1\hbar\left[\mathcal{\hat H}_{\text{AL}},\rho\right]+i\left(\ddt{\rho}\right)_{\text{spont.\,emis.}}
\eea
As we will be focusing on the sub-Doppler regime, we assume
\bea\label{eq:adiabatic}
\vel\ll\Gamma/k
\eea
with $\vel$ being the velocity. The inequality holds for  $T\ll13$~mK for lithium.  This inequality allows us to replace the full time derivative in the l.h.s.~of (\ref{eq:OBE}) by a partial spatial derivative times the atomic velocity  
$$\dt\rightarrow\vel\dz$$\\
Using the notation $\Omega_i(z)=\Omega_i\,\text{cos}(z+\phi_i)$, and setting $\hbar=k=1$ from here on:

\small
\bea
\hspace*{-15pt}i\vel\ddz{\rho_{22}}&\hspace*{-2pt}=&\hspace*{-2pt}-2i\ot (z)\, \text{Im}(\rho_{23})+i\frac{\Gamma}{2}\rho_{33}\label{eq:OBE1}\\
\hspace*{-15pt}i\vel\ddz{\rho_{11}}&\hspace*{-2pt}=&\hspace*{-2pt}-2i\oo (z)\,\text{Im}(\rho_{13})+i\frac{\Gamma}{2}\rho_{33}\label{eq:OBE2}\\
\hspace*{-15pt}i\vel\ddz{\rho_{23}}&\hspace*{-2pt}=&\hspace*{-2pt}(\delta_2-i\frac{\Gamma}{2})\rho_{23}+\ot(z)\left(\rho_{33}-\rho_{22}\right)-\oo(z)\rho_{21}\label{eq:OBE3}\\
\hspace*{-15pt}i\vel\ddz{\rho_{13}}&\hspace*{-2pt}=&\hspace*{-2pt}(\delta_1-i\frac{\Gamma}{2})\rho_{13}+\oo(z)\left(\rho_{33}-\rho_{11}\right)-\ot(z)\rho_{12}\label{eq:OBE4}\\
\hspace*{-15pt}i\vel\ddz{\rho_{21}}&\hspace*{-2pt}=&\hspace*{-2pt}(\delta_2-\delta_1)\rho_{21}+\ot(z)\rho_{31}-\ot(z)\rho_{23}.\label{eq:OBE5}
\eea
\normalsize

\noindent The solution of these equations yields the expression of $\for(\vel)$ and $\Gamma'$.
This semi-classical model is valid only for velocities above the recoil velocity  $\vel_{\rm{rec}}=\hbar k/m$ (corresponding to a temperature $mv_{rec}/k_B $ of about 6~$\mu$K for Lithium).  Different theoretical studies \cite{zerussians, zerussians1,Grynberg:1994,drewsen1995,menotti1997,dunn2007} as well as experiments \cite{gupta1993,malossi2005} have been performed on such a $\Lambda$-configuration in standing waves or similar systems. However in our $^7$Li experiment, we have the situation in which the $\Lambda$ configuration is coupled to a gray molasses scheme which involves a different set of dark states. This fixes the laser light parameters to values that have not been studied before and motivates our theoretical exploration. Thus we will concentrate on the situation corresponding to the conditions of our experiment.

To solve the OBEs (\ref{eq:OBE1}-\ref{eq:OBE5}), we first introduce a perturbative approach that enables us to point out the relevant physical mechanisms. We further extend the analysis by an exact approach in terms of continued fractions.

\subsection{Perturbative approach}

In our perturbative approach we choose a Rabi frequency $\ot$ between $2 \Gamma$ and $4 \Gamma$ and $\oo\ll\Gamma,\,\ot,\delta_{2}$ as the ratio of the repumper to principal laser power is very small, typically $(\oo/\ot)^{2} \lesssim .03$, under our experimental conditions.
We further simplify the approach by considering only the in-phase situation $\phi=0$; any finite phase would lead to divergencies of the perturbative approach at the nodes of wave 1.
The validity of  these assumptions will be discussed in section \ref{subsec:contfrac}.

\begin{figure}[htbp]
\begin{center}
\centerline{\includegraphics[width=1.\columnwidth]{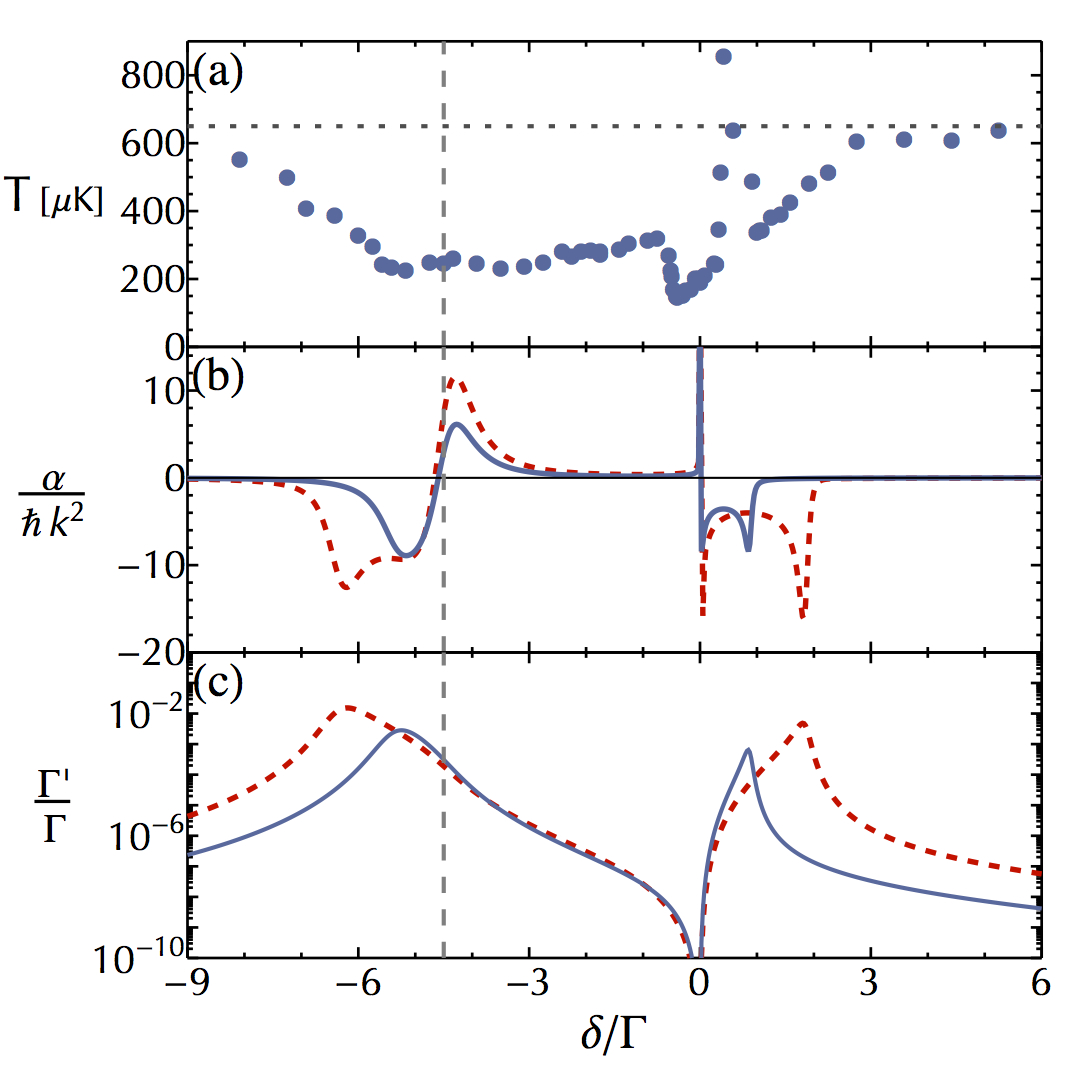}}
\caption{(Color online) Comparison of experimental data with the perturbative approach results for a detuning of the pump $\delta_2=2\pi\times 26.4\,{\rm MHz}=4.5\Gamma$,  (a) temperature versus repumper detuning, experiment, we indicate the MOT temparature by the dotted line, (b) and (c) show respectively the friction coefficient $\alpha$ and photon scattering rate $\Gamma'$ for $\ot=3.4\Gamma$, red dashed curve and $2.1\Gamma$, blue solid curve. The intensity ratio $(\oo/\ot)^{2}$ is 0.02. The vertical dashed line indicates the position of $\delta_1=0$.}
\label{fig:data1}
\end{center}
\end{figure}

We perform an expansion in powers of the Rabi frequency $\oo $ and the atomic velocity, such that the complete expansion reads:

\bea\label{eq:pertexp}
\rho_{ij}=\sum_{n,l}\rho_{i,j}^{(n,l)}(\oo)^n(\vel)^l.
\eea

This expansion of $\rho$ allows us to recursively solve the OBEs.  Using an expansion similar to eq.~(\ref{eq:pertexp}) for the force, we find:

\bea\label{eq:alpha}
\alpha=-\sum_{n=0}^{\infty} \for^{(n,1)}(\oo)^n.
\eea
We plug the perturbative solution of the OBEs into eq.~(\ref{eq:avgForce}) and find, to the lowest order ($n=2$) in $\oo$:
\bea
\alpha\simeq-\frac{(\oo)^2}{2\pi}\int_0^{2\pi}\text dz\,{\rm{sin}}(z) \left(\ot Re\,\rho_{23}^{(2,1)}+Re\,\rho_{13}^{(1,1)}\right)
\eea
The spontaneous emission rate to lowest order in $\vel$ and $\oo$ reads
\bea\label{eq:spont}
\Gamma'=\Gamma\,\frac{(\oo)^2}{2\pi}\int_0^{2\pi}\text dz\,\rho_{33}^{(2,0)}.
\eea

%


Figure~\ref{fig:data1} presents the results from (\ref{eq:alpha}, \ref{eq:spont}) compared with the experimental data.
It shows that indeed a narrow cooling force appears near the Raman resonance condition and that the photon scattering rate vanishes at exact resonance, hinting towards an increase of cooling efficiency with respect to the gray molasses Sisyphus cooling mechanism which achieves a temperature near 200\,$\mu$K over a broad range. The strong heating peak for small, positive repumper detuning is also a consequence of the negative value of $\alpha$, and the heating peak  shifts towards higher frequency and broadens for larger intensities of the principal laser. In contrast, the friction coefficient and scattering rate in the range $-6\leq \delta/\Gamma\leq -3$, which correspond to a repumper near resonance, do not seem to significantly affect the measured temperature.

To gain further physical insight into this cooling near the Raman condition, it is useful to work in the dressed-atom picture. Given the weak repumping intensity, we first ignore its effect and consider only the dressing of the states $\ket{2}$ and $\ket{3}$ by the strong pump with Rabi frequency $\ot$.  This dressing gives rise to an Autler-Townes doublet structure which follows the spatial modulation of the standing wave:
\bea\label{equ:atd}
\ket{2'}&\propto&\ket 2-i\ot(z)/\delta_2\ket 3\\
\ket{3'}&\propto&-i\ot(z)/\delta_2\ket 2+\ket 3
\eea

Since the pump is  relatively far detuned (in the conditions of Fig.\ref{fig:data1} $\ot/\delta_2\lesssim0.45$), the broad state $\ket{3'}$ carries little $\ket{2}$ character.  Conversely the narrow state $\ket{2'}$ is mostly state $\ket{2}$. It follows that $\ket{3'}$ has a lifetime  $\gamth\simeq \Gamma$ while $\ket{2'}$ is relatively long-lived with a spatially dependent linewidth $\gamt= \Gamma(\ot(z)/\delta_2)^2$ which is always $\leq \Gamma/6$ for the parameters chosen here. 
In order to reintroduce the effects of the repumping radiation, we note that the position in $\delta$ of the broad state is $\delth\simeq-\delta_2-\ot(z)^2/\delta_2$ and the narrow state $\delt\simeq\ot(z)^2/\delta_2$.
As coherent population transfer between $\ket{1}$ and $\ket{2'}$ does not change the ensemble temperature, we consider only events which couple atoms out of $\ket{2'}$ to $\ket 1$ through spontaneous decay and therefore scale with $\Gamma_{\ket{2'}}$.
The rates of coupling from $\ket 1$ into the dressed states can be approximated by the two-level absorption rates:
\small
\bea
\gamma_{\ket 1\rightarrow\ket{2'}}&\sim&\frac{\oo(z)^2}{2}\frac{\gamt(z)}{(\gamt(z)/2)^2+(\delta-\delt(z))^2}\label{eq:rate21}\\
\gamma_{\ket 1\rightarrow\ket{3'}}&\sim&\frac{\oo(z)^2}{2}\frac{\Gamma}{(\Gamma/2)^2+(\delta-\delth(z))^2}\label{eq:rate23}
\eea
\normalsize
Finally, these results are valid only in the limit $|\delta| > \Gamma \ot^2/\delta_2^2$ (see {\it e.g.}~\cite{arimondo1996}) when state $\ket{1}$ is weakly coupled to the radiative cascade. Near the Raman resonance, the dressed state family contains a dark state which bears an infinite lifetime under the assumptions made in this section but is in reality limited by off-resonant excitations and motional coupling. This dark state reads:
\bea
\ket{{\rm NC}}= (\ot\ket{1}-\oo\ket{2})/\sqrt{\oo^2+\ot^2},
\eea
\noindent which we must add in by hand.


\begin{figure}[!htbp]
\begin{centering}
\begin{overpic}[width=.8\columnwidth]{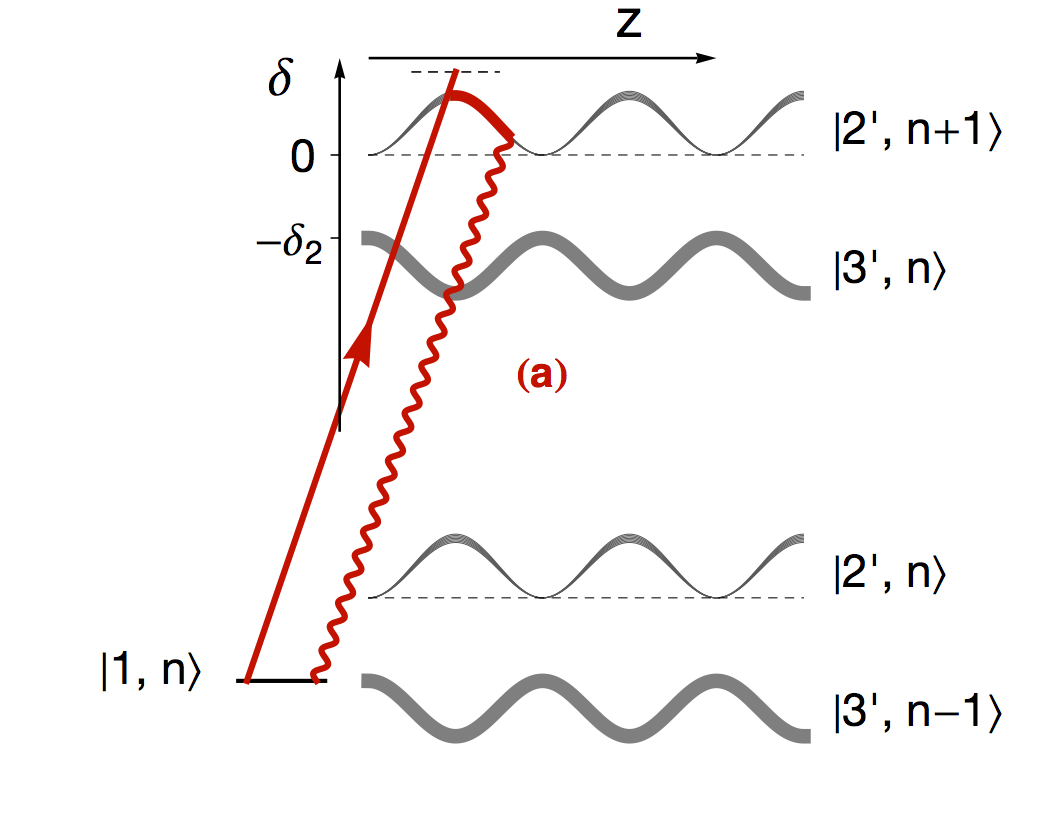}\end{overpic}
\begin{overpic}[width=.8\columnwidth]{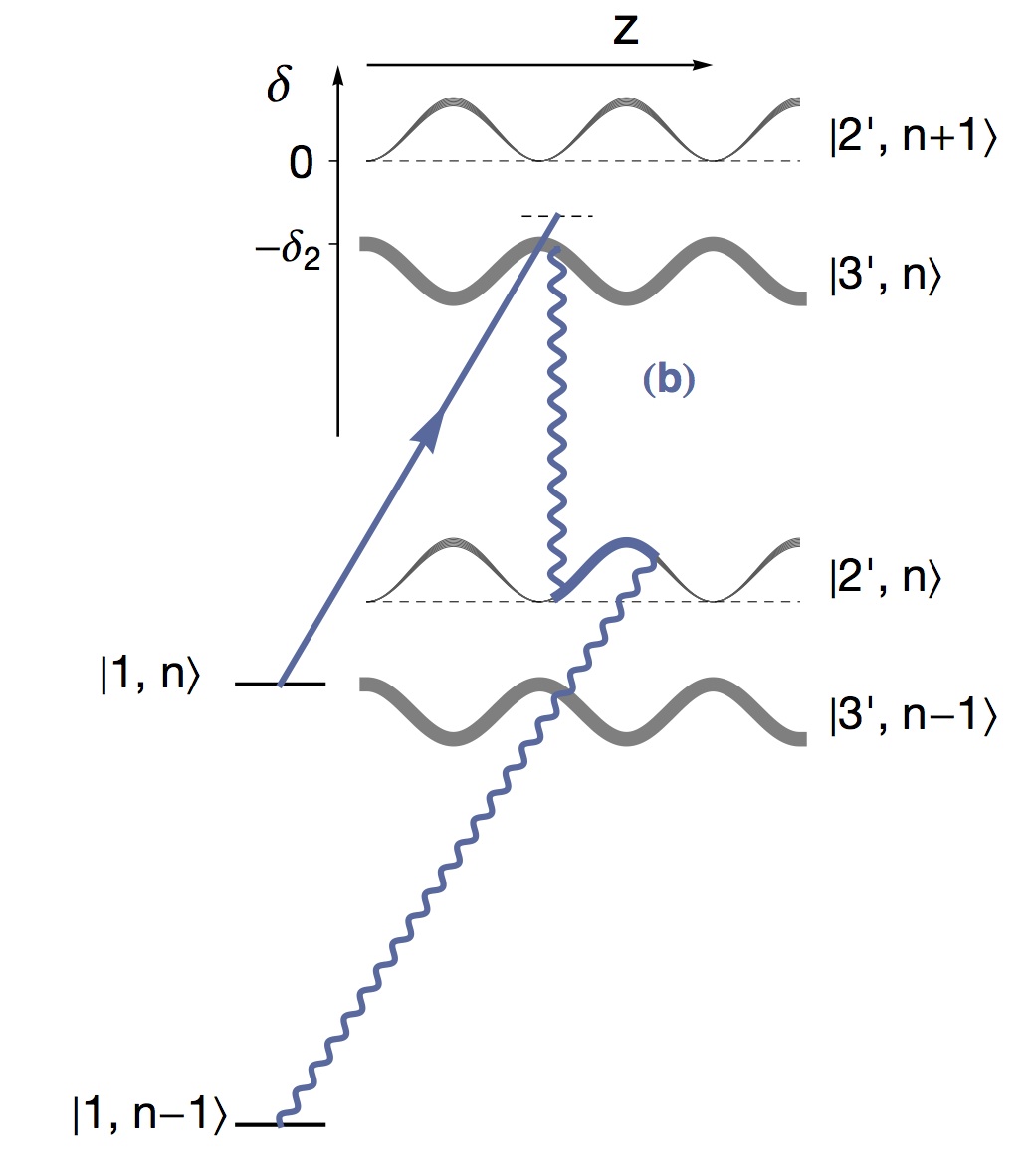}\end{overpic}\\
\caption{(Color online) The cascade of levels dressed by transition 2 with a schematical representation of state $\ket 1$. Traces show typical cycles of atoms pumped from $\ket 1$ and back depending on the detuning of wave 1. The detuning of the repumper modulates the entry point into the cascade of the dressed states, leading either a) heating or b) cooling processes.}
\label{fig:cascade}
\end{centering}
\end{figure}

Using this toy model, we now explain the features of Fig.~\ref{fig:data1} and Fig.~\ref{fig:data0}.  Figure~\ref{fig:cascade} represents the cascade of dressed levels where each doublet is separated by one pump photon. It gives rise, for example, to the well-known Mollow triplet.
Condition (\ref{eq:adiabatic}) states that if an atom falls in state $\ket{3'}$ it will rapidly decay to $\ket{2'}$ without traveling a significant distance.  However, the atom will remain in $\ket{2'}$ long enough to sample the spatial variation of the standing wave and gain or lose energy depending on the difference of light shift between the entry and departure points, as in most sub-Doppler cooling schemes. 


Let us first analyze the spontaneous emission rate shown in Fig.~\ref{fig:data1}(c). It reaches two maxima, the first one for $\delta\sim\delth$ and the second one for $\delta\sim\delt$, and it goes to exactly zero at $\delta=0$. The two maxima are simply due to scattering off the states $\ket{2'}$ and $\ket{3'}$. At $\delta=0$, $\Gamma'$ goes to zero due to coherent population trapping in $\ket{{\rm NC}}$.  It is the presence of this dark state which leads to the reduced scattering rate of photons around $\delta=0$ and the suppression of the final temperature of the gas in the region around the Raman condition.

The friction coefficient, Fig.~\ref{fig:data1}(b), displays a more complicated structure with variations in $\delta$. It shows a dispersive shape around $\delth$; remains positive in the range $\delth<\delta<0$; diverges at $\delta=0$; and reaches negative values for $\delta>0$ up to $\delt$, where it drops to negligible values. This structure for $\alpha$ can be explained using our toy model.  Let us consider the different scenarios corresponding to both sides of $\delta$ near 0, they follow formally from equations (\ref{eq:rate21},\ref{eq:rate23}) and the spatially varying linewidth of $\ket{2'}$.

For the case of the repumper tuned slightly blue of the narrow doublet state, $\delta>\delt$, shown in Fig.~\ref{fig:cascade}(a), the atoms are pumped directly from $\ket{1}$ into $\ket{2'}$. However, this pumping happens preferentially at the antinodes of the standing wave as the repumper intensity is greatest, the linewidth of $\ket{2'}$ is the largest, and the light shift minimizes the detuning of the repumper from the $\ket{1}\to\ket{2'}$ transition for the $\phi=0$ case considered here.  On average, the atoms exit this state at a point with a smaller light shift through a spontaneous emission process either into the cascade of dressed states or directly back to $\ket{1}$.  As a result, we expect heating and $\alpha<0$ in this region.


For repumper detunings between $\delth$ and $0$, Fig.~\ref{fig:cascade}(b), we predict cooling.  For this region, the atoms are initially pumped into $\ket{3'}$.  Here the light shift modifies the relative detuning, favoring coupling near the nodes of the light.  Spontaneous decay drops the atoms near to the nodes of the longer-lived $\ket{2'}$, and they travel up the potential hill into regions of larger light shift before decaying, yielding cooling and a positive $\alpha$.  These sign changes of $\alpha$ and the decreased scattering rate due to $\ket{\rm NC}$ in the vicinity of the Raman condition explain the features of our perturbative model.


We conclude this section by stating that the experimentally observed change of sign of the force close to the Raman condition is well described in our perturbative model. The model further reveals the importance of Raman coherence and the existence of a dark state. The dark state together with the friction coefficient associated with cycles represented in trace 5(b) correspond to a cooling mechanism analogous to that of gray molasses. In this way, the bichromatic system provides an additional gray molasses scheme involving both hyperfine states which complements the gray molasses cooling scheme on the principal transition. On the other hand, when the friction coefficient is negative in the vicinity of the two-photon resonance, it turns into a heating mechanism that overcomes the standard gray molasses operating on the $F=2\to F'=2$ transition.

The perturbative approach successfully revealed the mechanisms giving rise to the experimentally observed additional cooling. However, it also possesses some shortcomings. First, the divergence of $\alpha$ at $\delta=0$ is not physical; the assumption that $\oo$ is the smallest scale in the problem breaks down when $\delta\to 0$.  Alternatively, it can be seen as the failure of our model based on non-degenerate perturbative theory in the region where $\ket{1}$ and $\ket{2}$ become degenerate when dressed with $\omega_{1}$ and $\omega_{2}$, respectively.  Secondly, we have only addressed the $\phi=0$ case. Since the experiment was done in 3 dimensions with 3 pairs of counter-propagating beams, the relative phase between the two frequencies varies spatially, and we must test if the picture derived at $\phi=0$ holds when averaging over all phases.   In order to address these limitations and confirm the predictions of the perturbative approach, we now present a continued fractions solution to the OBEs which does not rely on $\oo$ being a small parameter.

\subsection{Continued fractions approach}\label{subsec:contfrac}

The limitations listed above can be addressed by using a more general approach, namely an expansion of the density matrix in Fourier harmonics:
\bea
\label{eq:fourierexp}
\rho_{ij}=\sum_{n=-\infty}^{n=+\infty}\rho_{ij}^{(n)}e^{i n k z}.
\eea

\noindent 
Injecting this expansion in (\ref{eq:OBE1}-\ref{eq:OBE5}) yields recursive relations between different Fourier components of $\rho$. Kozachiov \etal~\cite{zerussians,zerussians1} express the solutions of these relations for a generalized $\Lambda$ system in terms of continued fractions. Here we use their results to numerically solve the Bloch equations.  We then compute the force $\for(\vel)$ to arbitrary order of $\oo$ and extract $\alpha$ by means of a linear fit to the small-$\vel$ region.  We then compute $\for(\vel)$ and the photon scattering rate $\Gamma'$ averaged over the phase between the two standing waves.

\begin{figure}[htbp]
\begin{center}
\centerline{\includegraphics[width=.85\columnwidth]{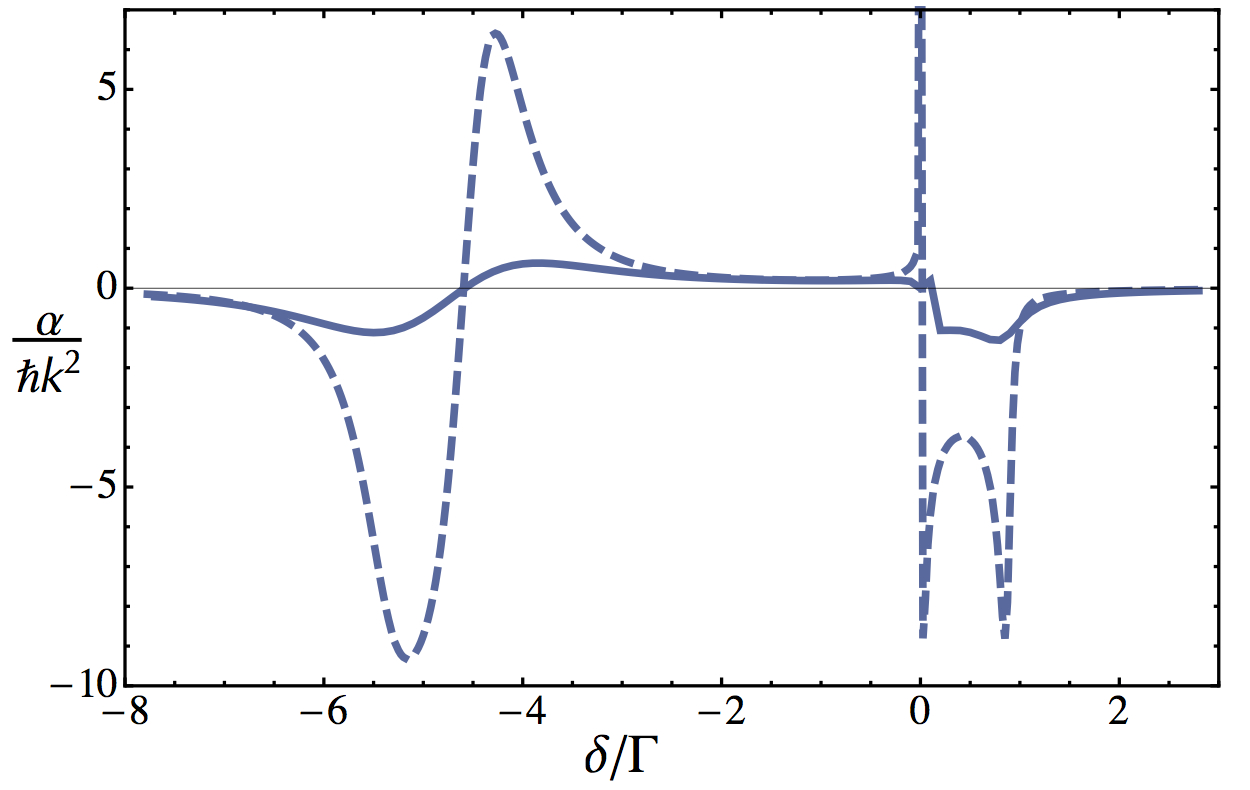}}
\caption{Comparison of results using the perturbative calculation (dashed), and the continued fractions (solid) for the $\phi=0$ case, with the same parameters as in Fig.~(\ref{fig:data1}) and $\ot=2.1\Gamma$.}
\label{fig:comp}
\end{center}
\end{figure}

Fig.~\ref{fig:comp} compares $\alpha(\delta)$ obtained through the continued fractions approach with the results of the perturbative expansion for the $\phi=0$ case. The continued fractions approach has removed the divergence at $\delta=0$ and $\alpha$ crosses zero linearly. The overall friction coefficient is reduced  but the two methods show qualitative agreement in the range of $\delta$ 
considered. At the Raman condition the interaction with light is canceled due to the presence of $\ket{\text{NC}}$, thus the diffusion coefficient $\mathcal D_{p}$ in momentum space also cancels. To lowest order, the diffusion and friction coefficients scale as
\bea
&\mathcal D_{p}\,\,\simeq& \delta^2\\\label{eq:difdelta}
&\alpha\,\,\simeq& \delta\label{eq:alphadelta}
\eea
according to (\ref{eq:limitT}) the temperature scales as: 
\bea
T \simeq \delta.
\eea
Through this qualitative scaling argument, we show that even though the light action on the atoms is suppressed when approaching the Raman condition, we expect that the temperature will drop when approaching from the $\delta<0$ side, completing the physical picture derived in the previous section.

Next, we analyze how a randomized phase between the repumping and principal standing waves, $\phi$, modifies $\for(\vel)$.  In order to take this into account, we calculate the phase-averaged force:
\bea
\left<\for(\vel)\right>_\phi=\frac{1}{2\pi}\int_0^{2\pi}\for(\vel,\phi)\,d\phi.
\eea

\begin{figure}[htbp]
\begin{center}
\centerline{\includegraphics[width=1.0\columnwidth]{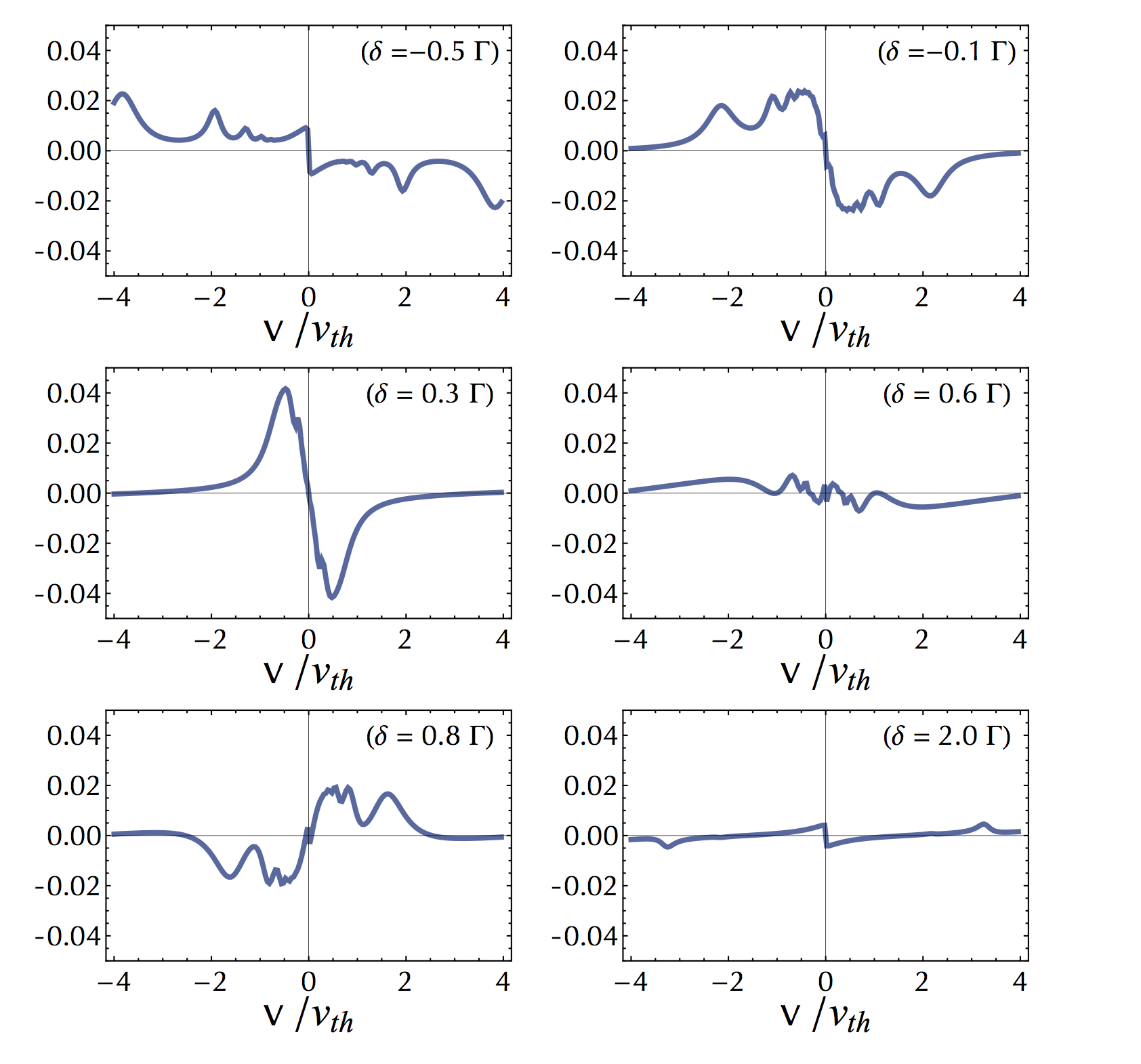}}
\caption{$\left<\for\right>_\phi$ in units of $1/\hbar k \Gamma$ as a function of $\vel$ for different values of $\delta$ around $\delta=0$. The horizontal scale is in units of the thermal velocity at $T=200\,\mu K$, $v_{th}=\sqrt{k_{{\rm B}} T/m}$.}
\label{fig:phaseaveraged}
\end{center}
\end{figure}
In Fig.~\ref{fig:phaseaveraged}, the phase-averaged force is plotted for various detunings near the Raman condition.
It can be seen that a cooling force is present for small detunings,  qualitatively in agreement with our perturbative model and with the experimental data. The force, however, changes sign to heating for small blue detuning, close to $\delta=0.6\,\Gamma$, also in qualitative agreement with the experimental data.  We note that the cooling slope very close to zero velocity in the $\delta=0.8\,\Gamma$ plot corresponds to a velocity on the order of or below the single-photon recoil velocity, \emph{i.e.} is non-physical.\\

\begin{figure}[htbp]
\begin{center}
\centerline{\includegraphics[width=0.95\columnwidth]{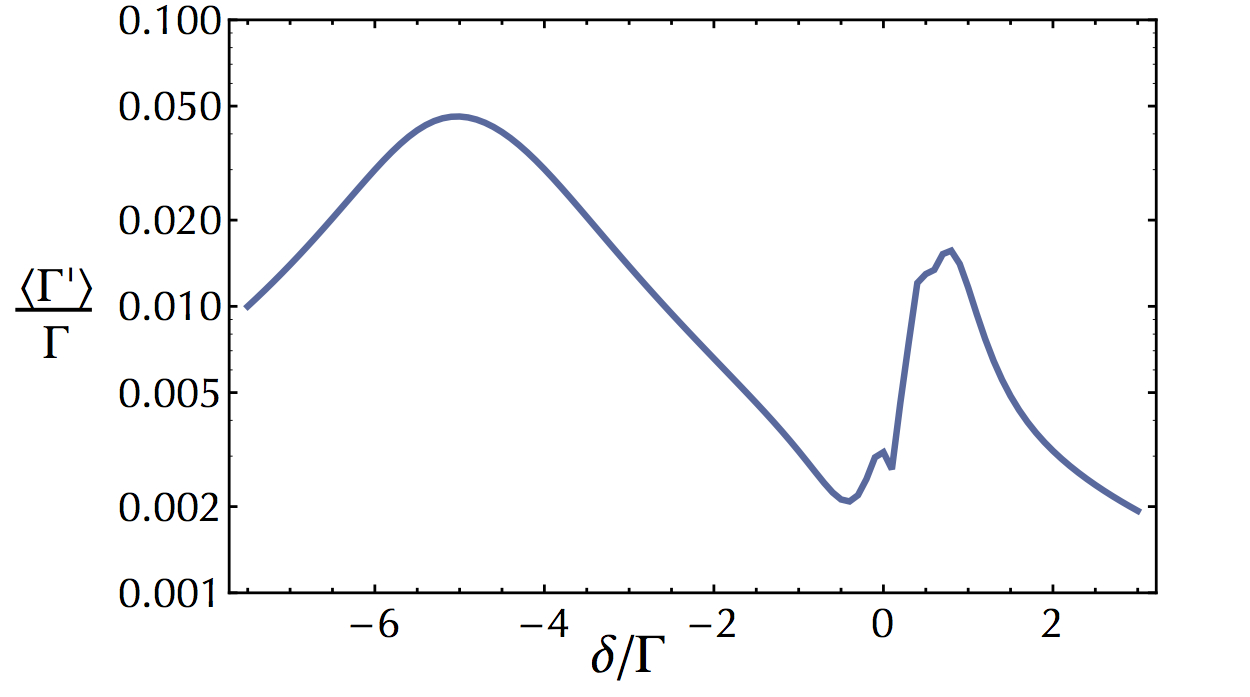}}
\caption{Continued fractions solution of the photon scattering rate $\Gamma ' = \Gamma\,\rho_{33}$ averaged over all relative phases of the repumper and principal standing waves as a function of the two-photon detuning $\delta$. Velocity-dependent effects are taken into account here by computing an average of $\left<\Gamma '\right>_\phi(\vel)$ weighed by a Maxwell-Boltzmann velocity distribution at 200 $\mu$K.}
\label{fig:phaseaveragedD}
\end{center}
\end{figure}

Finally, for the $\phi\neq0$ case, $\ket{{\rm NC}}$ varies in space and the motion of the atoms can couple atoms out of $\ket{{\rm NC}}$ even at the Raman condition.  In Fig.~\ref{fig:phaseaveragedD} we verify that the rate of photon scattering retains a minimum near the $\delta=0$ region after averaging over $\phi$ by plotting $\left<\Gamma '\right>_\phi=\Gamma \left< \rho_{33}\right>_\phi$ calculated with the continued fractions approach.  Overall, the friction coefficient $\alpha$ and photon scattering rate $\Gamma '$ confirm the existence of a cooling force associated with a decrease in photon scattering in the vicinity of the Raman condition for the 1-D bichromatic standing wave model.  Thus, the continued fractions calculation has confirmed the physical mechanisms revealed by the perturbative expansion and that the lowest temperatures should be expected close to $\delta=0$, as seen in the experiment.

\section{Conclusion}
In this study, using bichromatic laser light near  670-nm, we have demonstrated sub-Doppler cooling of $^7$Li atoms down to 60 $\mu$K with near unity capture efficiency from a magneto-optical trap. Solving the optical Bloch equations for a simplified $\Lambda$ level structure, we have analyzed the detuning dependence of the cooling force and photon scattering rate. Our analysis shows  that the lowest temperatures are expected for a detuning of the repumping light near the Raman condition, in agreement with our measurements. There the $\Lambda$-configuration adds a new set of long-lived dark states that strongly enhance the cooling efficiency. For $^7$Li, this addition results in a threefold reduction of the steady-state temperature in comparison with an incoherently repumped gray molasses scheme.  This atomic cloud at $60\mu$K is an ideal starting point for direct loading into a dipole trap, where one of the broad Feshbach resonances in the lowest-energy states of $^{7}$Li or $^{6}$Li could be used to efficiently cool the atoms to quantum degeneracy \cite{HuletUV,Gross:2008bp}.  Alternatively, when the atoms are loaded into a quadrupole magnetic trap, we measure a phase space density of ${\simeq}\, 10^{-5}$.
This $\Lambda$-enhanced sub-Doppler-cooling in a $D_1$ gray molasses is general and should occur in all alkalis. Notably, we have observed its signature in a number of the alkali isotopes not amenable to polarization gradient cooling: $^7$Li (this work), $^{40}$K \cite{fernandes2012}, and $^6$Li \cite{sievers2013}.

\begin{acknowledgements}
We acknowledge fruitful discussions with Y. Castin, J. Dalibard S. Wu, F. Sievers, N. Kretzschmar, D. R. Fernandes, M. Schleier-Smith, and I. Leroux, and support from R\'egion \^Ile de France (IFRAF-C'Nano), EU (ERC advanced grant Ferlodim), Institut de France (Louis D. Foundation), and  Institut Universitaire de France.\\
\end{acknowledgements}

\bibliographystyle{apsrev}


\end{document}